\documentclass[11pt]{article}
\usepackage{amsfonts,amsmath,amssymb,graphicx}
\usepackage[totalwidth=385pt,totalheight=580pt]{geometry}
\newfont{\twelvemsb}{msbm10 scaled\magstep1}
\newfont{\eightmsb}{msbm8}
\newfam\msbfam
\textfont\msbfam=\twelvemsb \scriptfont\msbfam=\eightmsb
\catcode`\@=11
\def\Bbb{\ifmmode\let\next\Bbb@\else
\def\next{\errmessage{Use \string\Bbb\space only in math mode}}\fi\next}
\def\Bbb@#1{{\fam\msbfam{{#1}}}}

\newcommand{\resection}[1]{\setcounter{equation}{0}\section{#1}}

\newcommand{\be}{\begin{equation}}
\newcommand{\ee}{\end{equation}}
\newcommand{\ba}{\begin{eqnarray}}
\newcommand{\ea}{\end{eqnarray}}
\newcommand{\nn}{\nonumber}

\newcommand{\m}{\mathcal}

\newcommand{\te}{\theta}
\newcommand{\ka}{p}
\newcommand{\cc}{{\bf c}}
\newcommand{\eps}{\epsilon}
\newcommand{\di}{\text{dim}}

\def \parton#1{\left( #1 \right)}
\def \parqua#1{\left[ #1 \right]}

\begin{document}
\sloppy
\renewcommand{\thefootnote}{\fnsymbol{footnote}}
\renewcommand{\thefootnote}{\arabic{footnote}}
\setcounter{footnote}{0}

\newpage
\begin{titlepage}
\vskip 1.2cm
\begin{center}
{\Large{\bf Exact results  for the  low energy
$AdS_4\times \mathbb{CP}^3$ string theory }}
\end{center}
\vspace{1.8cm}

\centerline{{ \large Alessandro Fabbri$^a$,
Davide Fioravanti$^a$, Simone Piscaglia$^a$ and Roberto Tateo$^b$\footnote{e-mail: alessandro.fabbri@bo.infn.it, fioravanti@bo.infn.it, piscagli@bo.infn.it, tateo@to.infn.it.}}}
\vskip 0.9cm

\centerline{${}^a$ \small INFN-Bologna and Dipartimento di Fisica ed Astronomia,
Universit\`a di Bologna,
}
\centerline{\small Via Irnerio 46, 40126 Bologna, Italy.}
\vskip 0.2cm
\centerline{${}^{b}$ \small Dipartimento di Fisica  and INFN,
Universit\`a di Torino,}
\centerline{\small Via P.\ Giuria 1, 10125 Torino, Italy.}
\vskip 0.2cm

\vskip 1.25cm

\begin{abstract}
\noindent
We derive the  Thermodynamic Bethe Ansatz equations  for the relativistic 
sigma model describing the  $AdS_4\times \mathbb{CP}^3$ string II A 
theory at strong coupling ({\it i.e.} in the Alday-Maldacena decoupling  limit). 
The corresponding  $Y$-system  involves  an infinite number of $Y$ functions and is of a new type, although it shares a peculiar feature with the $Y$-system for $AdS_4\times \mathbb{CP}^3$. A  truncation of the 
equations at  level $\ka$ and a further generalisation to generic rank $N$
allow us an alternative description of the theory as
the $N=4$, $\ka = \infty$ representative in an infinite family of  models 
corresponding to the conformal cosets 
$(\mathbb{CP}^{N-1})_{\ka} \times U(1)$,  perturbed by a  relevant composite field
$\phi_{(N,\ka)} =\phi_{[(\mathbb{CP}^{N-1})_\ka]} \times  \phi_{[U(1)]}$ that couples the two
independent  conformal field theories.  The calculation of the ultraviolet 
central charge confirms  the conjecture by Basso and Rej and   
the conformal dimension of the perturbing operator, at  
every $N$ and $\ka$, is obtained using the Y-system periodicity. 
The conformal dimension
of $\phi_{[(\mathbb{CP}^{N-1})_\ka]}$  matches  that of the field identified     by Fendley
while discussing  integrability issues  for the  purely  bosonic $\mathbb{CP}^{N-1}$
sigma model. 
\end{abstract}
\vspace{2cm}
\end{titlepage}
%
%
\resection{Introduction}
The theory of quantum exactly solvable models is currently  playing an important role
in the study of the  gauge/gravity correspondence \cite{Malda,Witten} as  many powerful  integrable model   methods 
were recently adapted to investigate  perturbative and nonperturbative 
aspects  in  multicolor QCD  \cite{L}  and various  branches of the $AdS/CFT$ duality \cite{MZ}.

The purpose of this paper is to study, through the Thermodynamic Bethe Ansatz (TBA) \cite{zamTBA,klassen-melzer},
the finite-size corrections of the (integrable) two-dimensional 
$\mathbb{CP}^{N-1}$  quantum sigma model minimally coupled to a massless Dirac fermion plus a Thirring term, as described in \cite{basso-rej}. Despite the original $\mathbb{CP}^{N-1}$ model (without the fermion) has been intensively studied, helping physicists with its underlying phenomenology to understand the (irrelevant) r\^ole of instantons in the real QCD and 
sharing, with the latter 4d  theory, the property of confinement \cite{Witten:1978bc}, the system considered here has received much less attention. However, very recently it has been discovered~\cite{Bykov:2010tv} that the $N=4$ case describes the strong coupling limit of the planar $AdS_4\times \mathbb{CP}^3$ string IIA sigma model: this is the low energy Alday-Maldacena decoupling  limit, which has given rise to the $O(6)$ non-linear sigma model in the $AdS_5\times \mathbb{S}^5$ case \cite{Alday:2007mf}. In fact,  this relativistic $\mathbb{CP}^{3} \times U(1)$   sigma model  gives an 
effective (low energy) description of the Glubser, Klebanov and Polyakov (GKP) spinning string 
dual to composite operators in ${\cal N}=6$ supersymmetric Chern-Simons  
built with a pair of bi-fundamental matter fields plus an infinite sea of 
covariant derivatives  acting on them. For large t'Hooft coupling, the low-lying excitations over this vacuum are relativistic and precisely described by this massive sigma model with $SU(4)\times U(1)$ symmetry.

For general $N$, the Lagrangian  of  the $SU(N)\times U(1)$ symmetric  model  under consideration  is  \cite{basso-rej} ({\it cf.} also \cite{Bykov:2010tv} for $N=4$)
\be
{\cal L}= \kappa( \partial_{\mu}-i A_\mu) \bar{z}( \partial^{\mu}-i A^\mu) z
+ i \bar{\psi} \gamma^{\mu}(\partial_{\mu} - i k A_{\mu}) \psi - \frac{\lambda_T}{2} 
(\bar{\psi} \gamma_{\mu} \psi)^2,
\label{Lag}
\ee
where the bosonic multiplet $z=(z_1,\dots, z_N)$ satisfies the constraint $\bar{z} z=1$, $k$
is  the fermion charge (and equals $2$ in \cite{Bykov:2010tv} for $N=4$) and the Thirring coupling needs to be fine-tuned as $\lambda_T=-\frac{k^2}{2 N\kappa}$ (and equals $-\frac{1}{2\kappa}$ in \cite{Bykov:2010tv}).
 Many important aspects of the  model (\ref{Lag}) were recently discussed by Basso and Rej in 
\cite{basso-rej} and more recently in \cite{Basso:2013pxa}.
In the current paper  we shall  start from the asymptotic Bethe Ansatz equations
proposed in \cite{basso-rej}
 and derive the set of  TBA equations describing the exact finite-size corrections 
of  the vacuum energy on a cylinder. 
Although most of the results presented here  are rigorously derived only for  $N=4$
 it is  possible, 
just through simple considerations, to conjecture  equations for  general values of $N$. 
Furthermore, borrowing  the idea that  2d sigma models can be viewed as the infinite level 
limit  of  a sequence of  quantum-reduced  field theories  associated  to 
perturbed conformal field theories (CFT),  we introduce   a set of  TBA 
equations classified by a   pair of integer parameters: the rank  $N$ and the level $\ka$ of conformal coset models  
\be
(\mathbb{CP}^{N-1})_p \times U(1)=
\frac{SU(N)_{\ka}}{SU(N-1)_{\ka} \times U(1)} \times U(1),
\ee
or equivalently, through the level-rank duality, of the  systems
\be
 (W^{(\ka)})_{N} \times U(1) = \frac{SU(\ka)_{N-1} \times  SU(\ka)_{1}  }{SU(\ka)_{N}} 
\times U(1),
\ee
where $W^{(\ka)}$ denotes the $SU(\ka)$-related   family of $W$-algebra  minimal models.
The rest of this paper is organized as follows. In  Section \ref{sting}, 
starting from the  asymptotic 
Bethe Ansatz equations  for the fundamental excitations \cite{basso-rej}, we  formulate the string hypothesis 
and derive the  TBA equations.
The corresponding Y-systems and the TBA equations  in 
 Zamolodchikov's universal form,  for the whole family 
of quantum-reduced  models,  
 are reported in  Section~\ref{YS}. The numerical and analytic checks 
on the ultraviolet and infrared behaviors of the systems,  together with the perturbed conformal field theory
interpretation,  are discussed in Section \ref{UVlim}. Section \ref{conclusions}
contains our conclusions. The relevant S-matrix elements and TBA kernels are reported in
 Appendix~\ref{App:kernels}. Finally, in Appendix \ref{App:folding} we show an interesting analogy between the $Y$-system diagrams of the $\mathbb{CP}^{3} \times U(1)$ and the $O(6)$ non-linear sigma models (which parallels that between the diagrams of their corresponding all couplings (energies) theories, {\it i.e.} the $AdS_4\times \mathbb{CP}^3$ and $AdS_5\times S^5$ string sigma models, respectively).

\resection{The string hypothesis and asymptotic BA equations}
\label{sting}
The starting point of the analysis are the Asymptotic Bethe Ansatz (ABA) equations   in the NS sector of 
the $SU(4) \times U(1)$ symmetric  model  proposed in  \cite{basso-rej} 
\be
\begin{split}
e^{-imL\sinh\theta_k}&=\prod_{j\neq k}^M \ S(\theta_k-\theta_j)
\prod_{j=1}^{\bar{M}} \ t_1(\theta_k-\bar{\theta}_j) \prod_{j=1}^{M_{1}}
\left(\frac{\theta_k-\lambda_j+\frac{i\pi}{4}}{\theta_k-\lambda_j-\frac{i\pi}{4}}\right), \\
1&=\prod_{j\neq k}^{M_1}\left(\frac{\lambda_k-\lambda_j+\frac{i\pi}{2}}{\lambda_k-\lambda_j-\frac{i\pi}{2}}\right)
\,\prod_{j=1}^{M_2}\left(\frac{\lambda_k-\mu_j-\frac{i\pi}{4}}{\lambda_k-\mu_j+\frac{i\pi}{4}}\right)
\,\prod_{j=1}^{M}\left(\frac{\lambda_k-\theta_j-\frac{i\pi}{4}}{\lambda_k-\theta_j+\frac{i\pi}{4}}\right), 		 \\
1&=\prod_{j\neq k}^{M_2}\left(\frac{\mu_k-\mu_j+\frac{i\pi}{2}}{\mu_k-\mu_j-\frac{i\pi}{2}}\right)
\,\prod_{j=1}^{M_1}\left(\frac{\mu_k-\lambda_j-\frac{i\pi}{4}}{\mu_k-\lambda_j+\frac{i\pi}{4}}\right)
\,\prod_{j=1}^{M_3}\left(\frac{\mu_k-\nu_j-\frac{i\pi}{4}}{\mu_k-\nu_j+\frac{i\pi}{4}}\right), 		 \\
1&=\prod_{j\neq k}^{M_3}\left(\frac{\nu_k-\nu_j+\frac{i\pi}{2}}{\nu_k-\nu_j-\frac{i\pi}{2}}\right)
\,\prod_{j=1}^{M_2}\left(\frac{\nu_k-\mu_j-\frac{i\pi}{4}}{\nu_k-\mu_j+\frac{i\pi}{4}}\right)
\,\prod_{j=1}^{\bar{M}}\left(\frac{\nu_k-\bar{\theta}_j-\frac{i\pi}{4}}{\nu_k-\bar{\theta}_j+\frac{i\pi}{4}}\right), 		 \\
e^{-imL\sinh\bar{\theta}_k}&=\prod_{j\neq k}^{\bar{M}} \ S(\bar{\theta}_k-\bar{\theta}_j)
\prod_{j=1}^{M} \ t_1(\bar{\theta}_k-\theta_j) \prod_{j=1}^{M_{3}}\left(\frac{\bar{\theta}_k-\nu_j+\frac{i\pi}{4}}{\bar{\theta}_k-\nu_j-\frac{i\pi}{4}}\right),
\label{ABA}
\end{split}
\ee
\normalsize
where, with respect to \cite{basso-rej}, we have chosen the twist factor  $q=1$,  and   redefined the magnonic rapidities as
\be
 \lambda_{k} = \frac{\pi}{2}u_{1,k},~~
 \mu_{k} =\frac{\pi}{2}u_{2,k},~~
 \nu_{k} =\frac{\pi}{2}u_{3,k}~. 
\ee
In (\ref{ABA}) $M$, $\bar{M}$ and $M_l$  with $l=1,2,3$ indicate the number of 
spinons, antispinons and flavour-$l$ magnons, respectively.
As $L\rightarrow \infty$, in the thermodynamic limit, the dominant  
contribution to the free energy  comes from  magnon excitations  arranging  
themselves into strings \cite{takahashi-suzuki} 
of form
\be
\begin{split}
\lambda_{k a}^{(l)} &= \lambda_{k}^{(l)}+\frac{i\pi}{4}(l+1-2 a),~~ (a=1,\dots ,l), \\
\mu_{k b}^{(m)} &=\mu_{k}^{(m)}+\frac{i\pi}{4}(m+1-2 b),~~(b=1,\dots ,m), \\
\nu_{k c}^{(n)} &=\nu_{k}^{(n)}+\frac{i\pi}{4}(n+1-2 c),~~(c=1,\dots ,n). 
\end{split}
\label{strings}
\ee
The product over the  strings (\ref{strings}) of the  ABA equations (\ref{ABA}) yield 
\be
\begin{split}
e^{-imL\sinh\theta_k} &=\prod_{j\neq k}^M \ S(\theta_k-\theta_j)
\prod_{j=1}^{\bar{M}} \ t_1(\theta_k-\bar{\theta}_j) \prod_{l=1}^{\infty}
\prod_{j=1}^{M^{(l)}}\left[S_{1,l}\left(\theta_k -\lambda_j^{(l)}\right)\right]^{-1}, \\
1&= \prod_{j=1}^{M} S_{l,1}\left(\lambda_k^{(l)}-\theta_j\right)\prod_{m=1}^{\infty}  
\prod_{j=1}^{M^{(m)}}S_{l,m}\left(\lambda_k^{(l)}-\mu_j^{(m)}\right)
\\ &\times \prod_{l'=1}^{\infty}\prod_{j=1}^{M^{(l')}}
\left[S_{l,l'+1}\left(\lambda_k^{(l)} -\lambda_j^{(l')}\right)\right]^{-1}
\left[S_{l, l'-1}\left(\lambda_k^{(l)}-\lambda_j^{(l')}\right)\right]^{-1},		 \\
1&=\prod_{m'=1}^{\infty}\prod_{j=1}^{M^{(m')}}\left[S_{m,m'+1}\left(\mu_k^{(m)} -\mu_j^{(m')}\right)\right]^{-1}
\left[S_{m, m'-1}\left(\mu_k^{(m)}-\mu_j^{(m')}\right)\right]^{-1} \\
&\times
\prod_{n=1}^{\infty}  \prod_{j=1}^{M^{(n)}}S_{m,n}\left(\mu_k^{(m)}-\nu_j^{(n)}\right)
\prod_{l=1}^{\infty}  \prod_{j=1}^{M^{(l)}}S_{m,l}\left(\mu_k^{(m)}-\lambda_j^{(l)}\right),		\\
1&= \prod_{j=1}^{\bar{M}} S_{n,1}\left(\nu_k^{(n)}-\bar{\theta}_j\right)\prod_{m=1}^{\infty}  
\prod_{j=1}^{M^{(m)}}S_{n,m}\left(\nu_k^{(n)}-\mu_j^{(m)}\right) \\
&\times 
\prod_{n'=1}^{\infty}\prod_{j=1}^{M^{(n')}}\left[S_{n,n'+1}\left(\nu_k^{(n)} -\nu_j^{(n')}\right)\right]^{-1}
\left[S_{n, n'-1}\left(\nu_k^{(n)}-\nu_j^{(n')}\right)\right]^{-1},		 \\
e^{-imL\sinh\bar{\theta}_k} &=\prod_{j\neq k}^{\bar{M}} \ S(\bar{\theta}_k-\bar{\theta}_j)
\prod_{j=1}^{M} \ t_1(\bar{\theta}_k-\theta_j) \prod_{n=1}^{\infty}\prod_{j=1}^{M^{(n)}}\left[S_{1,n}\left(\bar{\theta}_k -\nu_j^{(l)}\right)\right]^{-1},
\end{split}
\label{BA2}
\ee
where $M^{(q)}$ is the number of length-$q$ strings, and we have introduced 
the  scattering amplitudes
\be
S_{l,m}(\theta)=\prod_{a=\frac{|l-m|+1}{2}}^{\frac{l+m-1}{2}}
\left(\frac{\theta-i\frac{\pi a}{2}}{\theta+i\frac{\pi a}{2}}\right)=
\prod_{a=1}^l
\left(\frac{\theta-\frac{i\pi}{4}(l+m+1-2 a)}{\theta+\frac{i\pi}{4}(l+m+1-2 a)}\right).
\ee
In this limit equations (\ref{BA2}) become
\be
\begin{split}
\sigma(\theta) &=m\cosh\theta+\m{K}*\rho(\theta)
+G*\bar{\rho}(\theta) - \sum_{l=1}^{\infty} K_{1,l}*\rho_l^{(1)}(\theta),\\
\sigma_n^{(1)}(\theta)&= K_{n,1}*\rho(\theta)+
\sum_{l=1}^{\infty} \left( K_{n,l}*\rho_l^{(2)}(\theta)- (K_{n,l+1}+K_{n,l-1})*\rho_{l}^{(1)}(\theta) \right), \\
\sigma_n^{(2)}(\theta) &=\sum_{l=1}^{\infty} \left(K_{n,l}*\rho_l^{(3)}(\theta) 
+K_{n,l}*\rho_l^{(1)}(\theta)-  
(K_{n,l+1}+K_{n,l-1})*\rho_{l}^{(2)}(\theta) \right), \\
\sigma_n^{(3)}(\theta) &=  K_{n,1}*\bar{\rho}(\theta)+\sum_{l=1}^{\infty} \left( 
K_{n,l}*\rho_l^{(2)}(\theta) -
(K_{n,l+1}+K_{n,l-1})*\rho_{l}^{(3)}(\theta) \right), 	\\
\bar{\sigma}(\theta) &=m\cosh\theta+ \m{K}*\bar{\rho}(\theta)
+G*\rho(\theta)
-\sum_{l=1}^{\infty}  K_{1,l}*\rho_l^{(3)}(\theta),																\\
\end{split}
\ee
where $n=1,2,\dots$ and   we have introduced the densities of accessible states for
spinons $\sigma$, antispinons $\bar{\sigma}$, for 
magnonic strings $\sigma_n^{(1)}$, $\sigma_n^{(2)}$, $\sigma_n^{(3)}$,
likewise the occupied state densities $\rho$, $\bar{\rho}$, $\rho_n^{(1)}$, $\rho_n^{(2)}$, $\rho_n^{(3)}$; 
the convolution operation $\ast$ has been defined as 
$f\ast g(\te)=\displaystyle\int_{-\infty}^{+\infty}\,f(\te-\te')\,g(\te')d\te'$.
Further, the kernels $\mathcal{K}(\te)$, $G(\te)$ and $K_{l,m}(\te)$ are listed and described in Appendix \ref{App:kernels}. 

At temperature $T=1/R$, setting
\be
\frac{\rho(\te)}{\sigma(\te)-\rho(\te)} =e^{-\eps_0(\te)}~,
~~\frac{\bar{\rho}(\te)}{\bar{\sigma}(\te) -\bar{\rho}(\te)} = 
e^{-\bar{\eps}_0(\te)}~,~~\frac{\rho_m^{(i)}(\te)}{\sigma_m^{(i)}(\te)-
\rho_m^{(i)}(\te)} = e^{-\eps_{(i,m)}(\te)}, 
\ee
and
\be
L_0(\theta) =\ln\left(1+e^{-\eps_0(\theta)}\right)~,~~\bar{L}_0(\theta)
=\ln\left(1+e^{-\bar{\eps}_0(\theta)}\right),~~L_{(i,m)}(\theta) = \ln \parton{1 + e^{- \eps_{(i,m)}(\theta)}},~~ 
\ee
with $i,m = 1,2,\dots$   the  following set of  
TBA equations are recovered:
\be
\begin{split}
\eps_0(\theta) &= i \alpha + mR\cosh\theta-\m{K}*L_0(\theta) 
- G*\bar{L}_0(\theta) -
\sum_{l=1}^{\infty}  K_{1,l}*L_{(1,l)}(\theta), \\
\eps_{(1,n)}(\theta) &=  K_{n,1}*L_0(\theta) 
-\sum_{l=1}^{\infty} \left(K_{n,l}*L_{(2,l)}(\theta)-  (K_{n,l+1}+K_{n,l-1})*L_{(1,l)}(\theta)\right), \\
\eps_{(2,n)}(\theta) &=\sum_{l=1}^{\infty} \left( (K_{n,l+1}+K_{n,l-1})*L_{(2,l)}(\theta) 
-  K_{n,l}*L_{(1,l)}(\theta) - K_{n,l}*L_{(3,l)}(\theta) \right), \\
\eps_{(3,n)}(\theta) &=K_{n,1}*\bar{L}_0(\theta) 
-\sum_{l=1}^{\infty} \left( K_{n,l}*L_{(2,m)}(\theta) -  (K_{n,l+1}+K_{n,l-1})*L_{(3,l)}(\theta) \right), \\
\bar{\eps}_0(\theta)& =-i \alpha  + R\cosh\theta-\m{K}*\bar{L}_0(\theta) 
-G*L_0(\theta) -\sum_{l=1}^{\infty}  K_{1,l}*L_{(3,l)}(\theta)~.
\label{TBAS}
\end{split}
\ee
In  (\ref{TBAS}), we have included  the chemical potential \cite{klassen-melzer}. $\lambda= e^{i \alpha}=1$ for the 
ground state,
 while $\lambda=e^{i \alpha}= -1$ corresponds
to  the first excited state \cite{Martins:1991hw,Fendley:1991xn} associated to the  lifting, 
due to tunnelling \cite{Tunneling},
of a two-fold vacuum  degeneracy  of the model \cite{basso-rej}.     
The expression for the $\alpha$-vacuum energy is
\be
\label{free-en}
E_{\lambda}(m,R)=-\frac{m}{2\pi}\int_{-\infty}^{\infty} \,d\theta\,\cosh\theta\,(L_0(\theta)+\bar{L}_0(\theta))~.
\ee
In the far infrared $R m \gg 1 $ region
\be
E_{\pm 1}(m,R) \simeq  \mp  \frac{2 m}{\pi} C_{(4,\infty)} K_1(m R),  
\ee
where $K_1(x)$ is the modified Bessel function. The coefficient $C_{(4,\infty)}$ will be directly  obtained
from  the  TBA equations  in Section \ref{UVlim} 
and should match the  number of  $SU(4)$ flavours: $C_{(4,\infty)}=4$, in agreement with \cite{basso-rej}.  

\resection{The Y-system and the TBA in universal form}
\label{YS}

\begin{figure}[ht]
\centering
\includegraphics[width=7cm]{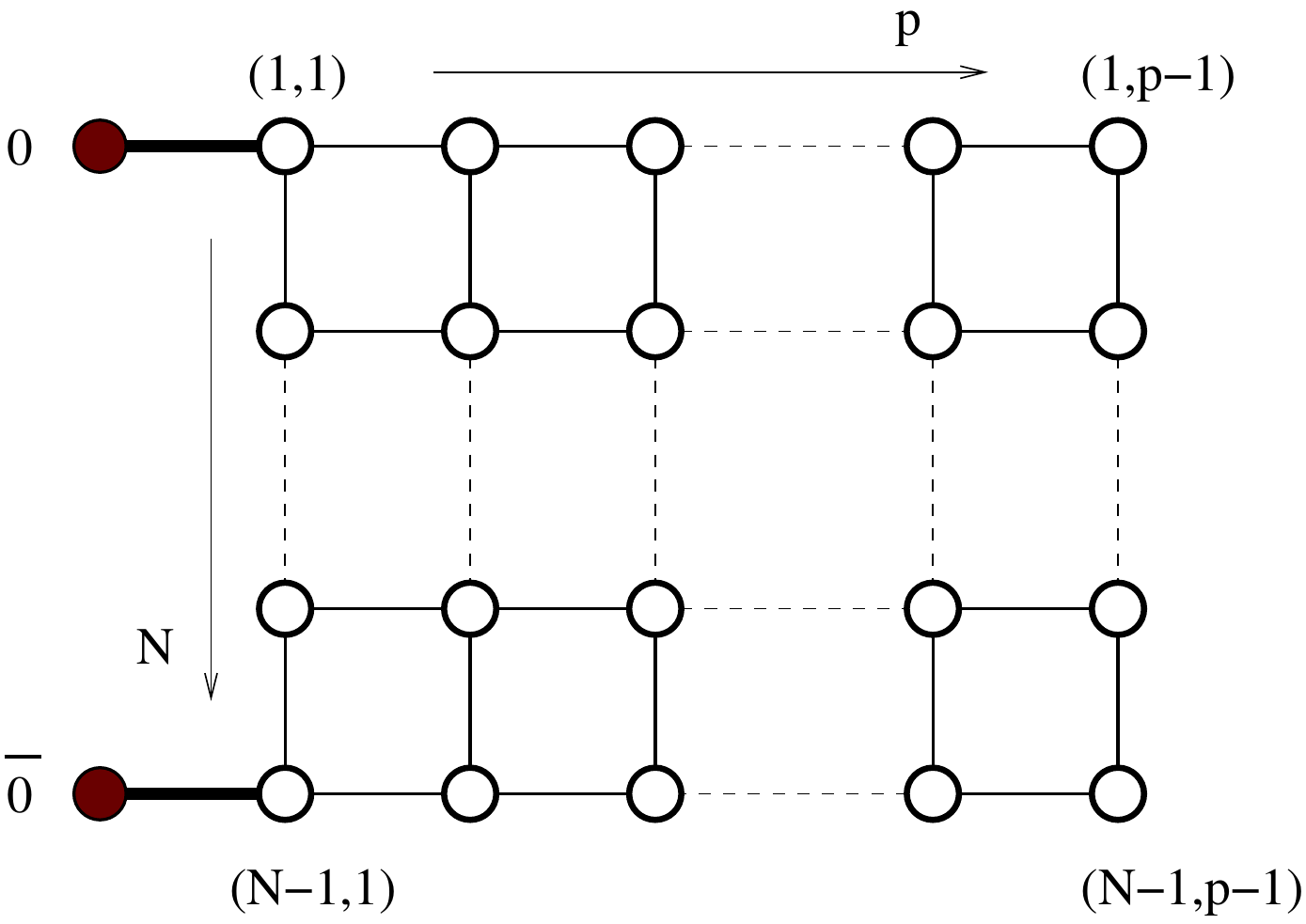}
\caption{ The $(\mathbb{CP}^{N-1})_{\ka} \times U(1)$ diagram.}
\label{RSS}
\end{figure}

Thanks to simple  identities for the TBA kernels \cite{zamUniversal,dynkinTBA}, the integral system (\ref{TBAS}) imply into the following  functional equations, the Y-system:
\be
\begin{split}
\label{YSS0}
Y_0(\te + i \frac{\pi}{2})  \, Y_0(\te - i \frac{\pi}{2})   &=  e^{-i 4 \alpha}
\frac{\bar{Y}_0(\te)}{Y_0(\te)} \parton{1 + Y_{(1,1)}(\te + i \frac{\pi}{4})} \parton{1 + Y_{(1,1)}(\te - i \frac{\pi}{4})} \parton{1 + Y_{(2,1)}(\te)}, \\
\bar{Y}_0(\te + i \frac{\pi}{2})\, \bar{Y}_0(\te - i \frac{\pi}{2})   &=  e^{i 4 \alpha}
\frac{Y_0(\te)}{\bar{Y}_0(\te)} \parton{1 + Y_{(3,1)}(\te + i \frac{\pi}{4})} \parton{1 + Y_{(3,1)}(\te - i \frac{\pi}{4})} 
\parton{1 + Y_{(2,1)}(\te)},
\end{split}
\ee
and, for the magnonic equations,
\be
\begin{split}
\label{YSS}
&Y_{(1,\,l)}(\te+i \frac{\pi}{4})\, Y_{(1,\,l)}(\te-i \frac{\pi}{4})  = \Big ( 1 + \delta_{l1}Y_0(\te) \Big) \frac{ \parton{1 + Y_{(1,\,l-1)}}(\te) \parton{1 + Y_{(1,\,l+1)}(\te)} } { \parton{1 + \displaystyle 
\frac{1}{Y_{(2,\,l)}}(\te) } } \\
&Y_{(2,\,l)}(\te+i \frac{\pi}{4}) \, Y_{(2,\,l)}(\te-i \frac{\pi}{4}) =  \frac{ \parton{1 + Y_{(2,\,l-1)}(\te)} \parton{1 + Y_{(2,\,l+1)}(\te)} } { \parton{1 + \displaystyle \frac{1}{Y_{(1,\,l)}(\te)} } \parton{1 + \displaystyle \frac{1}{Y_{(3,\,l)}(\te)} }} \\
&Y_{(3,\,l)}(\te+i \frac{\pi}{4}) \, Y_{(3,\,l)}(\te-i \frac{\pi}{4}) = \Big ( 1 + \delta_{l1}\bar{Y}_0 \Big) \frac{ \parton{1 + Y_{(3,\,l-1)}(\te)} \parton{1 + Y_{(3,\,l+1)}(\te)} } { \parton{1 + \displaystyle \frac{1}{Y_{(2,\,l)}(\te)} } }    ,
\end{split}
\ee
where  the Y functions are related to 
the pseudoenergies $\eps_A(\te)$, through
\begin{equation}
\label{Y-def}
Y_0(\theta) = e^{-\eps_0(\theta)},~~\bar{Y}_0(\theta) = e^{-\bar{\eps}_0(\theta)},~~
Y_{(i,l)}(\theta) = e^{\eps_{(i,l)}(\theta)}~. 
\end{equation}
Notice that the  RHS of (\ref{YSS0}), due to presence of the 
factor $\bar{Y}_0/Y_0$, does not have the 
standard  Y-system form  \cite{zamUniversal}. However, a more careful  inspection 
of the  TBA equations  reveals the presence of an important relation: 
\be
\frac{Y_0(\te + i \frac{\pi}{4})\, Y_0(\te - i \frac{\pi}{4})}{\bar{Y}_0(\te + i 
\frac{\pi}{4}) \, \bar{Y}_0(\te - i \frac{\pi}{4})}= e^{-i 4 \alpha} \frac{1+Y_{(1,1)}(\te)}{1+Y_{(3,1)}(\te)}.
\label{key-rel}
\ee
Using this in  (\ref{YSS0}) allows us to  recast the Y-system  into the following more standard-looking form
\be
\begin{split}
\label{YSS4}
Y_0(\te + i \frac{\pi}{2}) \, \bar{Y}_0(\te - i \frac{\pi}{2}) &= 
\left(1+Y_{(1,1)}(\te + i \frac{\pi}{4}) \right)\left(1+Y_{(2,1)}(\te) \right)\left(1+Y_{(3,1)}(\te - i \frac{\pi}{4}) \right), \\
\bar{Y}_0(\te + i \frac{\pi}{2}) \, Y_0(\te - i \frac{\pi}{2})  &=  
\left(1+Y_{(3,1)}(\te + i \frac{\pi}{4}) \right)\left(1+Y_{(2,1)}(\te) \right)\left(1+Y_{(1,1)}(\te - i \frac{\pi}{4}) \right),
\end{split}
\ee
together with the magnonic equations (\ref{YSS}).
Due to the appearance of the  mixed product $Y_0 \bar{Y}_0$ on the LHS of (\ref{YSS4}),  the latter equations are still 
slightly different from the systems discussed in the early  literature  on Y-systems~\cite{zamUniversal,dynkinTBA,Kuniba:1993cn}, while the the magnonic equations (\ref{YSS}) are rather standard. Therefore, the entire $Y$-system and subsequent universal TBA (see below) can be thought of as encoded in the diagram in {\it Fig.}(\ref{RSS}) with some caveats on the massive nodes (\ref{YSS4}).  
This novel type  of ``crossed'' Y-system,  without shifts on the RHS \footnote{Pictorially, the bold link between the massive node $0$ (${\bar 0}$) and the magnonic one in {\it Fig.}(\ref{RSS}) means that the shift in the LHS is twice that in the RHS, so that we need somehow to compensate and shift also the lower index, along the entire first (magnon) column. A similar bold link may be imagined in the case of the $O(2n)$ non-linear sigma model $Y-$system, in particular for $2n=6$ ({\it cf.} Appendix \ref{App:folding}).}, was  first 
obtained  in \cite{ads4} and \cite{Gromov:2009at}, in the context of the TBA for  anomalous dimensions in the  planar ${\cal N}= 6$ superconformal Chern-Simons, {i.e.} $AdS_4/CFT_3$. Pictorially, the related $Y$-system diagram \cite{ads4, Gromov:2009at} may be obtained from that for planar $AdS_5/CFT_4$ by means of some sort of 'folding' process of the two wings with doubling of the fixed row of massive nodes; the same relation seems to hold (at strong coupling) between their low energy decoupled models, namely the present $\mathbb{CP}^{3} \times U(1)$ \cite{Bykov:2010tv} and the $O(6)$ nonlinear sigma models \cite{Alday:2007mf}, respectively. We shall give some details on this issue in Appendix \ref{App:folding}. At last but not least, an intriguing example of ``crossed'' $Y$-system describes the strong coupling behaviour of the gluon scattering amplitudes in $SYM_4$ \cite{alda-sever-vie}.

Before concluding this section, we would like to make a final relevant generalisation. It is natural to consider a more general family of systems, stemming from the introduction of two positive integers $N$ and $\ka$, so that we conjecture for the massive nodes the equations
\be
\begin{split}
\label{YSSN}
Y_0(\te + i \frac{\pi}{2}) \, \bar{Y}_0(\te - i \frac{\pi}{2}) &= 
\prod_{l=1}^{N-1} \left(1+Y_{(l,1)}(\te + i \frac{\pi}{2}- i \frac{\pi l}{N}) \right), \\
Y_0(\te - i \frac{\pi}{2}) \, \bar{Y}_0(\te + i \frac{\pi}{2}) &=  
\prod_{l=1}^{N-1} \left(1+Y_{(l,1)}(\te - i \frac{\pi}{2}+ i \frac{\pi l}{N}) \right),
\end{split}
\ee
while for the magnonic nodes the relations
\ba
Y_{(i,j)}(\te + i \frac{\pi}{N}) \, Y_{(i,j)}(\te - i \frac{\pi}{N})  &=& 
\left( 1 + \delta_{i,1} \delta_{j,1} Y_0(\te)+ \delta_{i,N-1}\delta_{j,1}   \bar{Y}_0(\te) \right)\times  \nonumber\\
&\times & \prod_{l=1}^{\ka-1} \parton{1 + Y_{(i,l)}(\te)}^{A_{l,j}^{(\ka-1)}} 
 \prod_{l'=1}^{N-1} \parton{1 + \displaystyle \frac{1}{Y_{(l',j)}(\te)} }^{-A_{l',i}^{(N-1)}} \ \ ; 
\ea
obviously, the system studied so far is recovered by fixing $(N,\ka)$ to $(4,\infty)$. With this simple generalisation, we are able to describe    a previously-unknown infinite  family of Y-systems  
naturally associated to a generic  $SU(N)$ algebra    
with  quantum reduced coset level $\ka$.
As we shall see in the following section,  the    obtained   truncated family of 
Y-systems exhibit all  the 
important  features common to  more standard types of  Y-systems. In particular, 
they can be interpreted as  periodic sets  of  discrete recursion relations~
\cite{zamUniversal} and 
their  solutions lead to sum-rules~\cite{kirillov}  and  functional identities 
for the Rogers dilogarithm~\cite{Gliozzi:1994cs} (See equation (\ref{func})).

Although the reader should keep in mind that most of the results presented in this paper
have been  rigorously derived only for $(N,\ka)=(4,\infty)$, from now on we shall leave the two positive
integers $N$ and $\ka$  unconstrained. 
For later purpose, it is convenient to transform
the   Y-system   into the Zamolodchikov's universal TBA form \cite{zamUniversal}.
Thanks to  the  Fourier integrals in (\ref{FTI}), we obtain
\be
\begin{split}
\eps_0(\theta) &+ \bar{\eps}_0(\theta)=2 mR \cosh\theta - \sum_{l=1}^{N-1} 
\chi_{(1 -\frac{2l}{N})}*\Lambda_{(l,1)}(\te),\\
\eps_0(\theta)&- \bar{\eps}_0(\theta)= i 2 \alpha 
- \sum_{l=1}^{N-1} \psi_{(1 -\frac{2l}{N})}*\Lambda_{(l,1)}(\te), \\
\eps_{(i,j)}(\theta) &= \delta_{i,1}\delta_{j,1}  \phi_{\frac{N}{2}}*L_0(\te)+
\delta_{i,N-1} \delta_{j,1} \phi_{\frac{N}{2}}*\bar{L}_0(\te)+ \sum_{l=1}^{\ka-1} 
 A_{l,j}^{(\ka-1)} \phi_{\frac{N}{2}}*\Lambda_{(i,l)}(\te) -
 \sum_{l=1}^{N-1}  A_{l,i}^{(N-1)}\phi_{\frac{N}{2}}* L_{(l,j)}(\te),
\end{split}
\label{TBAU}
\ee
with $\alpha \in \{0,\pi\}$,
$\Lambda_A(\te)= \ln(1+e^{\eps_A(\te)})$ and  the  $\alpha$-vacuum energy
given by equation (\ref{free-en}) with 
\be
E_{\pm 1}(m,R) \simeq  \mp  \frac{2 m}{\pi} C_{(N,\ka)} K_1(m R),  
\ee
in the $R m \gg 1 $ infrared region. The coefficient $C_{(N,\ka)}$, which contains  information on the $SU(N)$-related 
vacuum  structure of  the  model at  $(N, \ka)$ generic \cite{Zamolodchikov:1991vh, Dorey:1996he},
will be determined in the following Section.

\resection{The ultraviolet and infrared limits}
\label{UVlim}
The models under consideration  can be thought of as  2d conformal field theories
 perturbed by a relevant operator  which becomes marginally relevant in the limit 
 $\ka \rightarrow \infty$ and 
whose vacuum energy  is given by the expression (\ref{free-en}) endowed with the groundstate TBA solution. 
In particular, the CFT is characterized by the value of its conformal anomaly, $\cc_{(N,\ka)}$,
which peculiarly enters the ($\alpha=0$) vacuum energy (\ref{free-en}) in the 
$mR \ll 1$ ultraviolet regime \cite{cardy}:
\be
E_{+1}(m,R) \simeq -\frac{\pi \cc_{(N,\ka)} }{6 R}. 
\ee
Thus, to obtain the central charge we have to study analytically the TBA equations in the limit  
$r=mR\rightarrow 0$. In this  limit the solutions $\eps_A(\te)$ to (\ref{TBAU}) 
develop a central plateau which broadens as $r$ approaches  zero~\cite{zamTBA,klassen-melzer}. 
The Casimir coefficient $\cc_{(N,\ka)}$ acquires contributions from  right and left kink-like 
regions, separately~\cite{zamTBA},
and the  result can be written as a sum-rule for the   Rogers dilogarithm function
\be 
\mathcal{L}(x) =-\frac{1}{2} \int_0^x \parqua{ \frac{\ln(1-t)}{t} + \frac{\ln t}{1-t}} 
dt,~~(0<x<1).
\ee
The final result is
\be
\cc_{(N,\ka)} = \cc^{(0)}_{(N,\ka)} - \cc^{(\infty)}_{(N,\ka)},
\ee
with
\be
\cc^{(0)}_{(N,\ka)} = 
\frac{6}{\pi^2} \parqua{ \mathcal{L} \parton{ \frac{y_0}{1+y_0} } +
 \mathcal{L} \parton{ \frac{\bar{y}_0}{1+\bar{y}_0} }
 + \sum_{i=1}^{N-1} \sum_{l=1}^{\ka-1} \mathcal{L} \parton{ \frac{y_{(i,l)}}{1+y_{(i,l)}}}},
\label{sum0}
 \ee
and
\be
\cc^{(\infty)}_{(N,\ka)}  = \frac{6}{\pi^2} \sum_{i=1}^{N-1} \sum_{l=1}^{\ka-1} \mathcal{L} \parton{ \frac{z_{(i,l)}}{1+z_{(i,l)}} }. 
\ee
The constants $y$s are given by the $\theta$-independent ({\it i.e.} stationary) solutions of the 
Y-system, while the $z$s are the stationary solutions of  (\ref{YSS}) with $Y_0=\bar{Y}_0=0$.  
The two relevant systems of  stationary equations are 
\be
\begin{split}
\label{YSSNS}
y_0 \bar{y}_0  &= \prod_{l'=1}^{N-1} \left(1+y_{(l',1)} \right),\\
(y_{(i,j)})^2  &= \left( 1 + \delta_{i,1} \delta_{j,1} y_0+ \delta_{i,N-1}\delta_{j,1}   \bar{y}_0 \right) 
\prod_{l=1}^{\ka-1} \parton{1 + y_{(i,l)}}^{A_{l,j}^{(\ka-1)}}  
\prod_{l'=1}^{N-1} \parton{1 + \displaystyle \frac{1}{y_{(l',j)}} }^{-A_{l',i}^{(N-1)}},
\end{split}
\ee
with $y_0=\bar{y}_0$ and  $y_{(i,j)}=y_{(N-i,j)}$  ($i=1,\dots,N-1$, $j=1,2,\dots$), and 
\be 
\label{plateau-ir-y}
(z_{(i,j)})^2  =  \prod_{l=1}^{\ka-1} \parton{1 + z_{(i,l)}}^{A_{l,j}^{(\ka-1)}}  
\prod_{l'=1}^{N-1} \parton{1 + \displaystyle \frac{1}{z_{(l',j)}} }^{-A_{l',i}^{(N-1)}}~.
\ee  
Finding the exact solutions to  equations (\ref{YSS},\ref{YSSNS}) for general $N > 3$ and $\ka$ 
turned  out to be  much  more difficult 
then expected. Setting $\varphi=\pi/(2(\ka+N-1))$, the results for lower ranks are the following\\
\begin{itemize}
\item $N=2$:
\be
y_{(1,i)}= (\ka-i)(\ka-i+2),~ y_{0}=\bar{y}_{0}=\ka,
\ee
with $i=1,2,\dots,\ka-1$. \\
\item $N=3$:
\be
y_{(1,i)}=y_{(2,i)}=\frac{\sin((\ka-i) \varphi) \sin((\ka-i+3)\varphi)}{\sin(\varphi) \sin(2\varphi) },
\ee
with  $i=0,1,\dots, \ka-1$ and $y_{0}=\bar{y}_{0}=y_{(1,0)}=y_{(2,0)}$.\\
\item  $N=4$:
\be
y_{(1,\ka-1)}=y_{(3,\ka-1)}= \frac{ 2 \sin(2\varphi)+\sin(6\varphi)+\sin(10 \varphi)}{2 \sin(6\varphi)},~
y_{(2,\ka-1)}=\frac{ 2 \sin(2\varphi)+\sin(6\varphi)+3 \sin(10 \varphi)}{ 2 \sin(2\varphi)+3 \sin(6\varphi)+\sin(10 \varphi)}.
\ee
(The   stationary values   for the remaining  Y functions 
can be  obtained using  (\ref{YSS}) and (\ref{YSSNS}) recursively.)
\end{itemize}

To deal with  the generic  $(N,\ka)$ case,  we  relied on  a high-precision numerical work 
to conjecture the exact result for the dilogarithm sum-rule (\ref{sum0}).
Starting from $\ka=2$  and $N=2$ we were able to obtain the  constants $y$s with a precision 
of about  $10^{-15}$, for $\ka < 20$ and $N<5$. 
The accuracy   progressively decreased down to 
$10^{-12}$ for  values around $\ka=61$ and $N=4$.
The numerical results  lead to  the following precise conjecture 
\be
\cc_{(N,\ka)}^{(0)} =  \frac{\ka (1+ \ka N- \ka)}{\ka+N-1}. 
\label{conj}
\ee
The constant $z$s are instead  analytically known to be~\cite{kirillov} 
\be
z_{(i,j)}=\frac{\sin((j+N) \phi) \sin(j \phi)}{ \sin((i+\ka) \phi) \sin(i \phi)},
\ee
with $\phi=\pi/(\ka+N)$, 
and the  corresponding  Rogers dilogarithm sum-rule is~\cite{kirillov}
\be
\cc_{(N,\ka)}^{(\infty)} = \frac{6}{\pi^2} \sum_{i=1}^{N-1} \sum_{l=1}^{\ka-1} 
\mathcal{L} \parton{ \frac{z_{(i,l)}}{1+z_{(i,l)}} } = \frac{\ka (N-1)(\ka-1)}{\ka +N}.
\label{ex1}
\ee
Finally, subtracting (\ref{ex1}) from (\ref{conj}) we obtain 
\be
\label{carica}
\cc_{(N,\ka)}= \frac{\ka(1-\ka-N +N^2 + 2 N \ka)}{(N+\ka)(N+\ka-1)} = \frac{\ka\; \di[SU(N)]}{ \ka+N} - 
\frac{\ka \; \di[SU(N-1)])}{ \ka+N-1} 
\ee
with $\di[SU(N)]= N^2-1$.  The  numerical outcome  for  the 
central charge at  $N=4$ for 
the  $\ka$-truncated models are
compared with equation (\ref{carica}) in Table \ref{T1}: the match is very good and   
  leaves little doubt on the correctness
of conjecture (\ref{conj}).
\begin{table}[tb]
\begin{center}
\begin{tabular}{|c|c|c|c|}
\hline
Level $\ka$  & Numerics &  Exact & Error \\
\hline
 2 & 1.8000000000000014 & 9/5    &  $1.3 \times 10^{-16}$ \\
 3 & 2.428571428571437  & 17/7   &  $8.4 \times 10^{-15}$ \\
 4 & 2.928571428571431  & 41/14   &  $2.6 \times 10^{-15}$ \\
 5 & 3.333333333333345  & 10/3    &  $6.7 \times 10^{-15}$ \\
 6 & 3.666666666666656  & 11/3    &  $1.1 \times 10^{-14}$ \\
 7 & 3.945454545454537  & 217/55  &  $8.4 \times 10^{-15}$ \\
 8 & 4.181818181818161  & 46/11   &  $2.0 \times 10^{-14}$ \\
 9 & 4.384615384615358  & 57/13   &  $2.7 \times 10^{-14}$ \\
10 & 4.56043956043953  & 415/91  &  $3.0 \times 10^{-14}$ \\
11 & 4.7142857142856   & 33/7    &  $1.1 \times 10^{-13}$ \\
41 & 6.212121212124    & 205/33  &  $2.8 \times 10^{-12}$ \\
51 & 6.35353535324     & 629/99  &  $2.9 \times 10^{-10}$ \\
61 & 6.4519230761      & 671/104 &  $8.2 \times 10^{-10}$ \\
\hline
\end{tabular}
\caption{\small $N=4$: comparison between numerics and equation (\ref{carica}).}
\label{T1}
\end{center}
\end{table}
In conclusion, the  central charge (\ref{carica}) deduced  from equations (\ref{YSS},\ref{YSSN}), 
coincides precisely with that   of the  coset model
\be
(\mathbb{CP}^{N-1})_p 
\times U(1)  = \frac{SU(N)_{\ka}}{SU(N-1)_{\ka} \times U(1)} \times U(1) 
\equiv \frac{SU(\ka)_{N-1} \times  SU(\ka)_{1}  }{SU(\ka)_{N}} \times U(1).
\label{id}
\ee 
The Casimir coefficient  for the $SU(N) \times U(1)$ sigma model is then recovered  in the 
limit $\ka \rightarrow \infty$:
\be
\cc_{(N,\infty)}=  \di[SU(N)]- \di[SU(N-1)]= 2 N-1.
\ee
Thus  $\cc_{(4,\infty)}=7$, a result that  coincides  with the value predicted in \cite{basso-rej} through a naive  degree of freedom counting  argument.

However, the identification of the model using only the Casimir coefficient is by no means  unique 
as, for example, the two $U(1)$ factors in (\ref{id}) yield compensating 
contributions to $\cc_{(N,\ka)}$ leading to an equivalently good   match with   
the central charge of the $\frac{SU(N)_{\ka}}{SU(N-1)_{\ka}}$ coset. 

To further support  the identification 
(\ref{id}), following \cite{zamUniversal}, we have  determined the conformal dimension $\Delta_{(N,\ka)}$
of  the perturbing operator using the intrinsic periodicity properties of the  
Y-system at finite $N$ and $\ka$.

Assuming arbitrary  initial conditions and using 
the Y-system as a recursion relation,  we descovered  that the following periodicity property holds
\be
Y_A \parton{ \theta + i \pi P_{(N,\ka)} } = Y_A(\theta),
\label{period}
\ee 
with  $P_{(N,\ka)}=\frac{2(\ka+N-1)}{N}$. Thus, according to \cite{zamUniversal} 
(cf. also \cite{Zam-RSOS,dynkinTBA}), we can  conclude that 
\be
\Delta_{(N,\ka)} = 1-\frac{1}{P_{(N,\ka)}}= 1 - \frac{N}{2(\ka+N-1)},
\label{del}
\ee
is  the conformal dimension of the  operator which perturbs the conformal field theory at 
finite $\ka$ and generic  $N$.  A first consequence of  (\ref{del}), is that the model
$\frac{SU(N)_{\ka}}{SU(N-1)_{\ka}}$ can be almost straightforwardly  discarded.  Furthermore, we have assumed that  the two  
CFTs, originally disconnected and respectively related   to  $(\mathbb{CP}^{N-1})_p$    
and   $U(1)$,   are tied together by the  perturbing operator 
$\phi_{(N,\ka)}$ in the simplest  possible   way:
\be
\phi_{(N,\ka)} =\phi_{[(\mathbb{CP}^{N-1})_\ka]} \times  \phi_{[U(1)]},~~
\Delta_{(N,\ka)}=\Delta_{[(\mathbb{CP}^{N-1})_\ka]}+ \Delta_{[U(1)]}.
\ee
For the identification of $\Delta_{[(\mathbb{CP}^{N-1})_\ka]}$ and $\Delta_{[U(1)]}$, the presence of two 
independent integer parameters was very important as both 
$\Delta_{[(\mathbb{CP}^{N-1})_\ka]}$ and  $\Delta_{[U(1)]}$  depend  nontrivially  on $N$ and 
$\ka$. At $\ka=1$, the TBA equations (\ref{TBAU}) reduce to those   for a free fermion. This fact
leads to
\be
\Delta_{[(\mathbb{CP}^{N-1})_1]}=0~,~~ \Delta_{[U(1)]}=\Delta_{(N,1)}=1/2. 
\label{con1}
\ee
At $N=2$, the TBA equations coincide with the   $D_{\ka+1}$ models with two  massive nodes and a
tail  of magnons. These groun dstate TBA equations  were  identified in \cite{Tateo:1994pb} 
(see, also \cite{dynkinTBA}) --up to possible orbifold ambiguities-- 
with  a particular series of points of the  fractional sine-Gordon model \cite{Bernard:1990ti}. 
The latter identification  leads to 
the further constant
\be
\Delta_{[(\mathbb{CP}^{1})_\ka]}= \frac{(\ka-1)}{\ka}~,~~
\Delta_{[U(1)]}= \frac{1}{\ka (\ka+1)}.
\label{con2}
\ee
Relations (\ref{con1}) and (\ref{con2}) together, allow to select the conformal dimension
uniquely:
\be
\Delta_{[(\mathbb{CP}^{N-1})_\ka]}= \frac{(\ka-1)(N+2 \ka)}{2 \ka (N+\ka-1)}~,~~
\Delta_{[U(1)]}= \frac{N}{2 \ka (N+\ka-1)}.
\ee
It is interesting to notice that for $\ka=2$ the dimension $\Delta_{[(\mathbb{CP}^{N-1})_\ka]}$
corresponds to the field $\phi_{21}$ of the $c<1$ minimal models ${\cal M}_{N+1,N+2}$, 
while for generic $N$ and $\ka$ it coincides precisely with the conformal dimension 
of the  field $(\ka,\bar{\ka},1)+(\bar{\ka},\ka,1)$ in the $W^{(\ka)}$ minimal model
$\frac{SU(\ka)_{N-1} \times  SU(\ka)_{1}}{SU(\ka)_{N}}$, mentioned   by Fendley 
\cite{fendleyCorto} while discussing  integrability issues related to   
the purely-bosonic  $\mathbb{CP}^{N-1}$ sigma model.

Finally,  following  \cite{Zamolodchikov:1991vh, Dorey:1996he} equations (\ref{TBAU}) furnish in the infrared regime $m R \gg 1$
\be
\eps_0(\te) -i \alpha  \simeq \bar{\eps}_0(\te) + i \alpha \simeq m R \cosh \te - \frac{1}{2} \sum_{l=1}^{N-1} 
\ln( 1 + z_{(l,1)}),
\ee
and consequently
\be
E_{\pm 1}(m,R) \simeq  \mp  \frac{2 m}{\pi} C_{(N,\ka)} K_1(m R),  
\ee
with
\be
C_{(N,\ka)}= \sqrt{\prod_{l=1}^{N-1} \left(1+z_{(l,1)}\right)}= \frac{\sin( N \phi)}{\sin(\phi)} \ ,
\label{limit}
\ee
where we defined $\phi=\pi/(N+p)$.
In the  sigma model limit $\ka \rightarrow \infty$, then $\phi \rightarrow 0$ and (\ref{limit}) gives   $C_{(N,\infty)}=N$, as expected.

\resection{Conclusions}
\label{conclusions}
In this paper we have proposed the Thermodynamic Bethe Ansatz equations and the Y-systems for an 
infinite  family of perturbed conformal field theories related to the  $\mathbb{CP}^{N-1}$ sigma 
models coupled to a massless Thirring fermion. 

Although the  main motivation of the  work was  the recently discovered description~\cite{Bykov:2010tv} of the low energy $AdS_4\times \mathbb{CP}^3$ string IIA sigma model  ((strong) decoupling Alday-Maldacena limit \cite{Alday:2007mf}), most of  the here derived results are of a much wider mathematical and physical interest. In particular, we have introduced   a novel family of periodic Y-systems classified in terms of a pair
of integers $(N,\ka)$. These functional relations differ from the standard Lie-algebra related ones, discussed for example in 
\cite{zamUniversal,dynkinTBA,Kuniba:1993cn}, in a  non trivial way. In fact, not only the {\it same} $Y$-function appears in each LHS of the massive node equations (\ref{YSS}), but the massive $Y$s appear in a ``crossed'' way ({\it cf.} also Appendix \ref{App:folding} for some considerations). 

Many important features  of Y-systems were  recently investigated and proved by means of very  powerful 
Cluster Algebra methods (see, for example the review \cite{Kuniba:2010ir}).
Within  the latter   mathematical setup, 
it would be important   to clarify whether the  Y-systems introduced   here are genuinely new objects or otherwise they  lead  to Cluster Algebra  quivers that are  mutation-equivalent to some of the known 
ABCD-related cases~\cite{Kuniba:2010ir} ({\it cf.}, for example, the discussion in Section 7.3 of \cite{Nakanishi:2010tt}). 

Some  of the  mathematical results presented here   correspond to   
numerical-supported conjectures and, although we  have little   doubt on their  
exact validity,  
it would   be still important   to prove them rigorously.  

The   main mathematical conjectures are:  
the Y-system periodicity (\ref{period}),  the stationary dilogarithm 
identities (\ref{conj}) and the following  non stationary sum-rules 
\be
\sum_{n=1}^{2(N+\ka -1)}  \left(  \mathcal{L}\left( \frac{\bar{Y}_{0}(n)}{1+ \bar{Y}_{0}(n)} \right)+ 
\mathcal{L}\left( \frac{Y_{0}(n)}{1+ Y_{0}(n)} \right)+ \sum_{i=1}^{N-1} \sum_{j=1}^{\ka-1} 
\mathcal{L}\left( \frac{Y_{(i,j)}(n)}{1+ Y_{(i,j)}(n)} \right) \right)= 2 \ka (1+\ka N -\ka)\frac{\pi^2}{6},
\label{func}
\ee
where
$ Y_A(n)= Y_A\left(\te+ i \frac{\pi}{N} n \right)$
are the solutions of the Y-system, obtained  recursively from (\ref{YSS}, \ref{YSSN})  with arbitrary initial conditions \cite{Gliozzi:1994cs}.

Concerning the  specific  $\mathbb{CP}^{3} \times U(1)$  sigma model, we have performed a non-trivial  computation of the 
ultraviolet central  charge from TBA/$Y$-system, confirming 
the results  predicted in \cite{basso-rej} through a naive counting of the degrees of freedom. In fact, our conclusions were reached using highly  non trivial dilogarithm identities and by  considering 
the  sigma model as the $\ka \rightarrow \infty$ representative  in the  family of  perturbed coset conformal field theories   $\frac{SU(4)_{\ka}}{SU(3)_{\ka} \times U(1)} \times U(1)$, and concerned also the perturbing field.

Apart from the physical and mathematical aspects mentioned above, there are many other  issues  that we  would like to address in the near future:   
the  kink vacuum structure, the exact S-matrix  and  the mass-coupling relation for the 
quantum truncated models, the numerical study of the TBA equations for
the excited states \cite{Bazhanov:1996aq} and the derivation of simpler 
non-linear integral equations for both  the 
groundstate  and the excited states \cite{Klummpe:1991vs}
are only a small 
sample of important  open problems  that deserve  further attention.

\medskip
\medskip
\noindent{\bf Acknowledgments--} 
We thank Diego Bombardelli, Andrea Cavagli\`a, Francesco Ravanini, Marco Rossi and  Gerard Watts for useful discussions and help. SP and AF are grateful respectively to the Centro de F\'isica do Porto and to IPhT-Saclay for kind hospitality. This project was  partially supported by INFN grants IS  FI11, P14, PI11, the Italian 
MIUR-PRIN contract 2009KHZKRX-007 {\it ``Symmetries of the Universe and of the Fundamental Interactions''}, 
the UniTo-SanPaolo research  grant Nr TO-Call3-2012-0088 {\it ``Modern Applications of String Theory'' (MAST)},
the ESF Network {\it ``Holographic methods for strongly coupled systems'' (HoloGrav)} (09-RNP-092 (PESC))
 and MPNS--COST Action MP1210 {\it ``The String Theory
Universe''}.
\appendix
\resection{Scattering amplitudes and TBA kernels}
\label{App:kernels}
This  appendix contains the  explicit expressions for scattering amplitudes and the corresponding  TBA kernels   used 
throughout the main  text.\\

\underline{\bf Spinon-Spinon scattering}\\

\noindent
The spinon-spinon $S$-matrix amplitude  \cite{basso-rej} is
\be
S(\theta) = - \frac{ \displaystyle  \Gamma \parton{ 1 + i \frac{\theta }{2 \pi} } 
\Gamma \parton{ \frac{1}{4} - i \frac{\theta }{2 \pi} } }{ \displaystyle  \Gamma \parton{ 1 - i \frac{\theta }{2 \pi}}
 \Gamma \parton{ \frac{1}{4} + i \frac{\theta }{2 \pi} }},
\ee
and the corresponding   kernel $\m{K}(\te)$
\be
\m{K}(\theta)=\frac{1}{2\pi i}\frac{\partial}{\partial\theta}\ln S(\theta),
\ee
which may be represented in several alternative  ways as
\footnote{It could be useful  to remind that 
\be
\psi(z) = \frac{\Gamma'(z)}{\Gamma(z)} = 
- \gamma_E - \sum_{n=0}^{\infty} \parton{ \frac{1}{z+n} - \frac{1}{n+1}}, 
\ee
where $\gamma_E$ stands for the Euler constant.}
\be
\begin{split}
\mathcal{K}(\theta) &= \frac{1}{4 \pi^2} 
\left(
 \psi \parton{ 1 + i \frac{\theta}{2 \pi} } + \psi \parton{ 1 - i \frac{\theta }{2 \pi}} - \psi \parton{ \frac{1}{4} + i \frac{\theta }{2 \pi} } 
- \psi \parton{ \frac{1}{4} - i \frac{\theta }{2 \pi} } 
\right)
 \\
&= \sum_{n=0}^{\infty} \parton{ \frac{1}{\pi} \frac{2\pi (n+1/4)}{\theta^2 + 
(2\pi (n+1/4))^2 } - \frac{1}{\pi} \frac{2\pi (n+1)}{\theta^2 + (2\pi (n+1))^2 }}\\
&= \int_{-\infty}^{\infty} \frac{d \omega}{2\pi} e^{ i \omega \theta} \, \frac{q- q^4}{1- q^4}, 
\end{split} 
\ee
with $q= \exp \parton{ - \frac{\pi}{2} |\omega|}$.
It is straightforward to get
\be
\int_{-\infty}^{\infty} d\theta \, \mathcal{K}(\theta) = \lim_{\omega \rightarrow 0} \hat{\mathcal{K}}(\omega) = 
\frac{3}{4}.
\ee\\

\underline{\bf Spinon-antispinon scattering} \\

\noindent
The $S$-matrix amplitude associated to the spinon-antispinon scattering is
\be
t_1(\theta) = 
\frac{\displaystyle \Gamma \parton{ \frac{1}{2} - i \frac{\theta }{2 \pi} } \Gamma \parton{ \frac{3}{4} + i \frac{\theta }{2 \pi} } }{ \displaystyle \Gamma \parton{ \frac{1}{2} + i \frac{\theta }{2 \pi}} 
\Gamma \parton{ \frac{3}{4} - i \frac{\theta }{2 \pi} } }.
\ee
Consequently the kernel $G(\theta)$ is
\be
G(\theta)=\frac{1}{2\pi i}\frac{\partial}{\partial\theta}\ln t_1(\theta),
\ee
explicitly
\be
\begin{split}
G(\theta) &= \frac{1}{4 \pi^2} \left( \psi \parton{ \frac{3}{4} + i \frac{\theta}{2 \pi} } + 
\psi \parton{ \frac{3}{4} - i \frac{\theta }{2 \pi}} - \psi \parton{ \frac{1}{2} + i \frac{\theta }{2 \pi} } - \psi \parton{ \frac{1}{2} - i \frac{\theta }{2 \pi} } \right) \\
&= \sum_{n=0}^{\infty} \parton{ \frac{1}{\pi} \frac{2\pi (n+1/2)}{\theta^2 + (2\pi (n+1/2))^2 } 
- \frac{1}{\pi} \frac{2\pi (n+3/4)}{\theta^2 + (2\pi (n+3/4))^2 }} \\
&= \int_{-\infty}^{\infty} \frac{d \omega}{2\pi} e^{ i \omega \theta} \, \frac{q^2 - q^3}{1- q^4}, 
\end{split} 
\ee
with  $q= \exp \parton{ - \frac{\pi}{2} |\omega|}$.
Then
\be
\int_{-\infty}^{\infty} d\theta \, G(\theta) = \lim_{\omega \rightarrow 0} \hat{G}(\omega) 
= \frac{1}{4}.
\ee\\

\underline{\bf Magnon bound state scattering}\\

\noindent
Magnonic string solutions scatter according to the amplitudes
\be
S_{l,m}(\theta) = \prod_{a = \frac{|l-m|+1}{2} }^{ \frac{l+m-1}{2} } 
\left(\frac{ \displaystyle \theta-i\frac{\pi a}{2}}{\displaystyle  \theta+i\frac{\pi a}{2}} \right),
\ee
from which 
\be
K_{l,m}(\theta)=\frac{1}{2\pi i}\frac{\partial}{\partial\theta}\ln S_{lm}(\theta) = 
\sum_{a = \frac{|l-m|+1}{2}}^{\frac{l+m -1}{2}} \frac{1}{\pi} \frac{a \pi / 2}{\theta^2 +
( a \pi /2)^2}.
\label{klm}
\ee
Fourier transforming (\ref{klm}) gives  
\be
\hat{K}_{l,m}(\omega) = \sum_{a = \frac{|l-m|+1}{2}}^{\frac{l+m -1}{2}} e^{-a |\omega|\pi/2} = \frac{e^{- 
\frac{|\omega|\pi}{4} |l-m|} - e^{- \frac{|\omega|\pi}{4} (l+m)}}{2 \sinh ( \pi|\omega|/4)},
\ee
and the matrix
\be
N_{l,m} = \int_{-\infty}^{\infty} d\theta K_{l,m}(\theta) = \hat{K}_{l,m}(0) = \min[l,m] = \frac{l+m - |l-m|}{2},
\ee
whose inverse is
\be
\hat{K}_{n,l}^{-1}(\omega) = 2 \cosh \parton{ \frac{|\omega|\pi}{4}} \delta_{nl} - \parton{\delta_{n,l-1} + \delta_{n,l+1}},
\ee
with
\be
\sum_l \hat{K}_{n,l}^{-1}(\omega) \hat{K}_{l,m}(\omega) = \delta_{n,m}.
\ee\\\

\underline{\bf Helpful Relations in Bootstrapping Matrices and Kernels}\\

Here we are reviewing the identities between scattering matrices (\emph{cfr} \cite{zamUniversal}\cite{dynkinTBA}) required in order to write
down the $Y$-system and universal form TBA
\be
  \begin{split}
&S_{lm}\left(\theta+\frac{i\pi}{4}\right)\,S_{lm}\left(\theta-\frac{i\pi}{4}\right)=
S_{l-1,m}\left(\theta\right)\,S_{l+1,m}\left(\theta\right)\,e^{2\pi i \Theta(\theta)\,\delta_{lm}}
\\
&t_1\left(\theta+\frac{i\pi}{4}\right)\, t_1\left(\theta-\frac{i\pi}{4}\right)=-S\left(\theta+\frac{i\pi}{4}\right)\, 
S\left(\theta-\frac{i\pi}{4}\right)\, [S_{11}(\theta)]^{-1}
\\
&S\left(\theta+\frac{i\pi}{2}\right)\,S\left(\theta-\frac{i\pi}{2}\right)
  =- \frac{t_1(\theta)}{S(\theta)} S_{12}(\theta) \,e^{2\pi i \Theta(\theta)}
\\
&t_1\left(\theta+\frac{i\pi}{2}\right)\,t_1\left(\theta-\frac{i\pi}{2}\right)=
-\frac{S(\theta)}{t_1(\theta)}
\\
&S_{lm}\left(\theta+\frac{i\pi}{2}\right)\,S_{lm}\left(\theta-\frac{i\pi}{2}\right)=
S_{l-2,m}\left(\theta\right)\,S_{l+2,m}\left(\theta\right)\,e^{2\pi i \Theta(\theta)\,I_{lm}}
  \end{split}
\ee
($\Theta(x)$ stands for the Heaviside step function, while $I_{lm}=\delta_{l-1,m}+\delta_{l+1,m}$ 
). These relations are reflected into the following ones, involving the kernels: 
\be\label{relazKernel}
\begin{split}
&K_{lm}\left(\theta+\frac{i\pi}{4}\right)+ K_{lm}\left(\theta-\frac{i\pi}{4}\right)=
K_{l-1,m}\left(\theta\right)+K_{l+1,m}\left(\theta\right)+\delta(\theta)\,\delta_{lm}
\\
&G\left(\theta+\frac{i\pi}{4}\right) + G\left(\theta-\frac{i\pi}{4}\right) 
=\m{K}\left(\theta+\frac{i\pi}{4}\right) + \m{K}\left(\theta-\frac{i\pi}{4}\right) - K_{11}(\theta)
\\
&\m{K}\left(\theta+\frac{i\pi}{2}\right)+\m{K}\left(\theta-\frac{i\pi}{2}\right)=
-\m{K}(\theta)+G(\theta)+K_{12}(\theta)+\delta(\theta)
\\
&G\left(\theta+\frac{i\pi}{2}\right)+G\left(\theta-\frac{i\pi}{2}\right)=\m{K}(\theta)-G(\theta)
\\
&K_{lm}\left(\theta+\frac{i\pi}{2}\right)+ K_{lm}\left(\theta-\frac{i\pi}{2}\right)=
K_{l-2,m}\left(\theta\right)+K_{l+2,m}\left(\theta\right)+\delta(\theta)\,I_{lm}+ 						\\
&\qquad\qquad+\delta_{l1}\,\delta_{m1}\left[\delta(\theta+\frac{i\pi}{4})+\delta(\theta-\frac{i\pi}{4})\right]
\end{split}
\ee
(the last relation makes sense
\footnote{Actually, the contact terms $\delta(\theta \pm i\frac{\pi}{4})$ are but a pretty
formal scripture: relations (\ref{relazKernel}) always appear in integrals and it is to be taken into account
a residue calculation, whose net result is equivalent to the effect of some kind of complex-argument defined delta function.} 
provided we define $K_{l,0} = 0 \ ,\ K_{l,-1} = -K_{l,1}$). Moreover, we find:
\be\label{relazKernel2}
\begin{split}
&\m{K}(\theta+\frac{i\pi}{2})+G(\theta-\frac{i\pi}{2})-K_{11}(\theta+\frac{i\pi}{4})\,=\,0
\\
&\m{K}(\theta-\frac{i\pi}{2})+G(\theta+\frac{i\pi}{2})-K_{11}(\theta-\frac{i\pi}{4})\,=\,0
\\
&\m{K}(\theta+\frac{i\pi}{2})+G(\theta-\frac{i\pi}{2})+K_{11}(\theta-\frac{i\pi}{4})\,=\,K_{12}(\theta)+\delta(\theta)
\\
&\m{K}(\theta-\frac{i\pi}{2})+G(\theta+\frac{i\pi}{2})+K_{11}(\theta+\frac{i\pi}{4})\,=\,K_{12}(\theta)+\delta(\theta)
\end{split}
\ee

\underline{\bf The universal kernels}\\

The kernels appearing in the Zamolodchikov's universal form of the TBA equations (\ref{TBAU}) are 

\be
\begin{split}
\int_{-\infty}^{\infty}\frac{d\omega}{2\pi}\frac{\cosh( \frac{\pi}{2} a \omega)}{\cosh( \frac{\pi \omega}{2})}\,e^{i \omega\theta} &=
\frac{2}{\pi} \frac{ \cos(a \pi/2) \cosh \te}{\cos(a \pi)+\cosh(2\theta)} =\chi_a(\te), \\
\int_{-\infty}^{\infty}\frac{d\omega}{2\pi}\frac{\sinh( \frac{\pi}{2} a \omega)}{\sinh( \frac{\pi \omega}{2})}\,e^{i \omega\theta} &=
\frac{1}{\pi} \frac{ \sin(a \pi)}{\cos(a \pi)+\cosh(2\theta)} = \psi_a(\te),\\
\int_{-\infty}^{\infty}\frac{d\omega}{2\pi}\frac{1}{2 \cosh( \frac{\pi \omega}{2 a})}
\,e^{i \omega\theta} &=\frac{a}{2 \pi\,\cosh(a \theta)}=\phi_a(\theta).
\label{FTI}
\end{split}
\ee

\resection{Folding diagrams}
\label{App:folding}

We wish now to discuss some features about a pictorial folding process of diagrams, by elucidating an inspiring resemblance between the $Y$-system diagrams for the $O(6)$ Non-Linear Sigma Model  and the $\mathbb{CP}^{3} \times U(1)$ model considered throughout this paper.
\newline

\begin{figure} [htbp]
\centering
\includegraphics[width=0.45\textwidth]{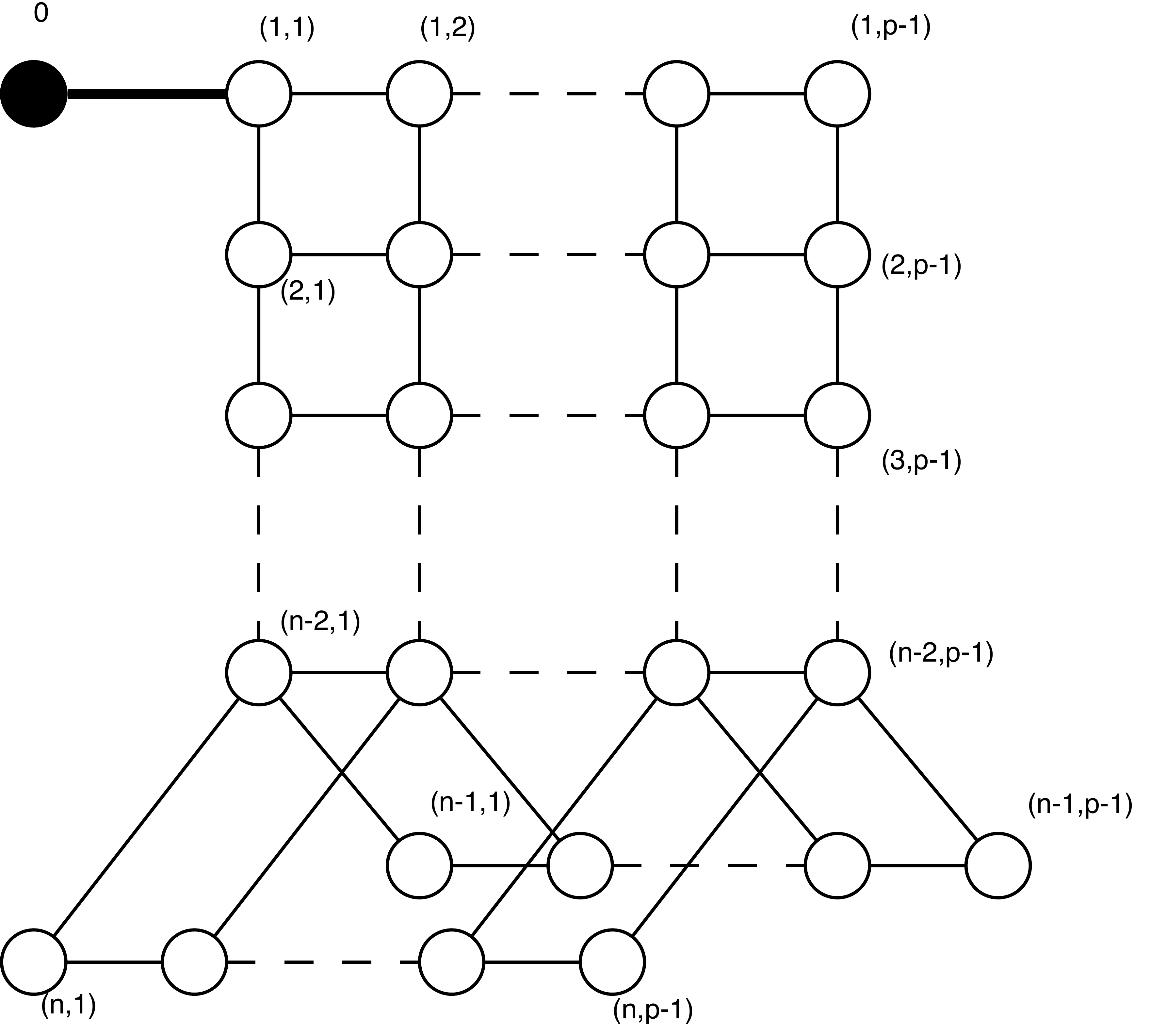}
\caption{The $O(2n)$ diagram. The labels of each node are associated to the functions $Y$ in (\ref{o(2r)ysys})}
\label{o(2r)diagram}
\end{figure}

\underline{\bf The $O(2n)$ Non-Linear Sigma Model TBA and Y-system}\\
According to \cite{fendleyCorto,BalHeg-virial,BalHeg-o(2r)} we can write the TBA system for the $O(2n)$ ($n\geq2$) Non-Linear Sigma Models as the limit of a certain sequence of coupled non-linear integral equations which read
\be
\label{o(2r)-tba-mass}
\begin{split}
&\epsilon_0(\theta) = m R \cosh \theta - \sum_{j=1}^{n-2} \chi_{\frac{2}{g}(n-1-j)}\ast L_{(j,1)}(\te)
-\phi_1 \ast [L_{(n-1,1)}+L_{(n,1)}]
\end{split}
\ee
\be
\label{o(2r)-tba-magn}
\begin{split}
\epsilon_{(a,m)}(\theta) = 
  -\delta_{m1}[\delta_{a1}+\delta_{a2}\delta_{n2}]\,\phi_{\frac{g}{2}}\ast L_0(\theta)
  -\,\phi_{\frac{g}{2}}\ast [L_{(a,m-1)}+L_{(a,m+1)}] +\sum_{b=1}^n\,I_{ab}\,\phi_{\frac{g}{2}}\ast \Lambda_{(b,m)}(\theta)
\end{split}
\ee
where $g=2(n-1)$ and $I_{ab}$ are respectively the Coxeter number and the incidence matrix associated to the $D_n$ Lie algebra,
while we defined
\be
L_0(\te) = \ln\parton{1+e^{-\epsilon_0(\theta)} } \qquad
L_{(a,m)}(\theta) = \ln\parton{1+e^{-\epsilon_{(a,m)}(\theta)} } \qquad
\Lambda_{(a,m)}(\theta) = \ln\parton{1+e^{\epsilon_{(a,m)}(\theta)} } \ \ . 
\ee
By means of the kernel relation
\small
\be
\chi_{\frac{2}{g}(n-1-j)} (\theta +\frac{i\pi}{2}) \, + \, \chi_{\frac{2}{g}(n-1-j)} (\theta -\frac{i\pi}{2})  =
\delta (\theta+\frac{i(n-1-j)\pi}{g})+ \delta (\theta-\frac{i(n-1-j)\pi}{g}) \,\,\, ,
\ee
and upon defining (as usual)
\be
\begin{split}
  &X_{(a\, ,m)}(\theta)= e^{-\epsilon_{(a,m)}(\theta)}\\
  &X_{0}(\theta)= e^{-\epsilon_{0}(\theta)}\qquad,
\end{split}
\ee 
equation (\ref{o(2r)-tba-mass}) entails
\be
\begin{split}
  \epsilon_0\parton{\theta+\frac{i\pi}{2}}+ & \epsilon_0\parton{\theta-\frac{i\pi}{2}}=
  -\sum_{a=1}^{n-2}\left[\ln\left(1+X_{(a,1)}(\theta-\frac{i(n-1-a)\pi}{g})\right)\right.+\\
  +& \left.\ln\left(1+X_{(a,1)}(\theta+\frac{i(n-1-a)\pi}{g})\right)\right]
  -\ln\left(1+X_{(n-1,1)}(\theta)\right)-\ln\left(1+X_{(n,1)}(\theta)\right) \qquad.
\end{split}
\ee
\normalsize
The latter is the first functional equation of the full $Y$-system \footnote{The only difference with respect to the $Y$-system derived in \cite{BalHeg-o(2r)} from the TBA \cite{fendleyCorto} is that we do not assume the symmetry (equality) between the two fork nodes $X_{(n,m)}$ and $X_{(n-1,m)}$.}
\be\label{o(2r)ysys}
\begin{split}
  X_0\parton{\theta+\frac{i\pi}{2}} & X_0\parton{\theta-\frac{i\pi}{2}}=
  \prod_{a=1}^{n-2} \left[\left(1+X_{(a,1)}(\theta-\frac{i(n-1-a)\pi}{g})\right)\right. \times\\
  &\times\left.\left(1+X_{(a,\,1)}(\theta+\frac{i(n-1-a)\pi}{g})\right)\right]  
  \left(1+X_{(n-1,\,1)}(\theta)\right)\left(1+X_{(n,\,1)}(\theta)\right)								 \\
  X_{(a,m)}\parton{\theta+\frac{i\pi}{g}} & X_{(a,m)}\parton{\theta-\frac{i\pi}{g}}=
  [1+\delta_{1m}(\delta_{a1}+\delta_{n2}\delta_{a2})X_0(\theta)]
  \frac{(1+X_{(a,m+1)}(\theta))(1+X_{(a,m-1)}(\theta))}{\displaystyle\prod_{b=1}^n \left(1+\frac{1}{X_{(b,m)}(\theta)}\right)^{I_{ab}}} \,\,\, ,
\end{split}
\ee
which may be encoded in the diagram of \textit{Fig.}(\ref{o(2r)diagram}) \footnote{This diagram and its interpretation is slightly different from those of \cite{BalHeg-o(2r)}.}. The bold link has the same meaning (explained in footnote 2 on page 6) as in the $\mathbb{CP}^{N-1} \times U(1)$ model diagram of \textit{Fig.}(\ref{RSS}). 

\begin{figure} [hbtp]
\centering
\includegraphics[width=0.45\textwidth]{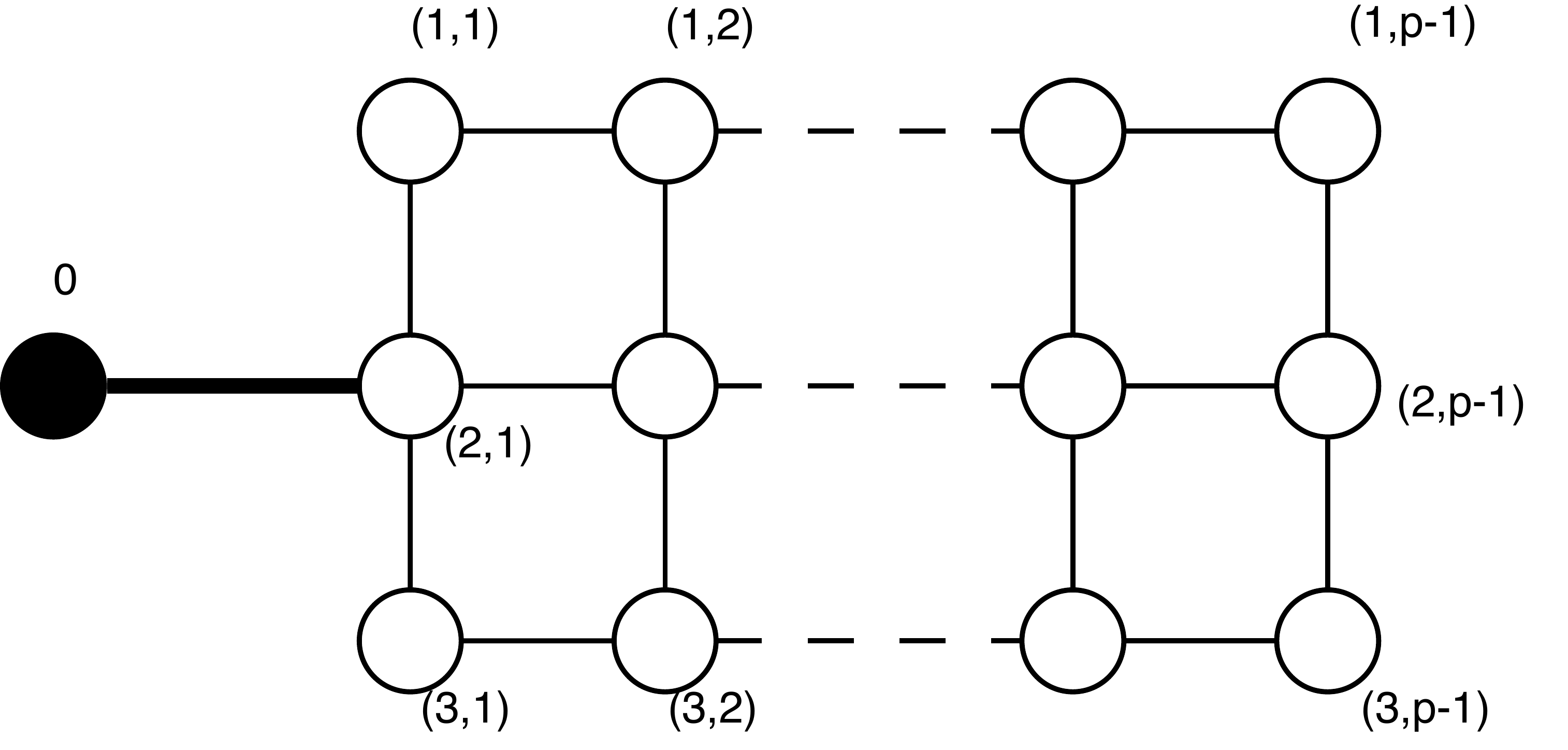}
\caption{The $O(6)$ diagram. The labels of each node are to be intended as the subscripts of the functions $X$ appearing in (\ref{o(6)ysys}).}
\label{o(6)diagram}
\end{figure}

\underline{\bf Folding diagrams}\\
In the particular case $n=3$, the $Y$-system of the $O(6)$ non-linear sigma model reads
\be
\begin{split}
X_0(\te+\frac{i\pi}{2}) X_0(\te-\frac{i\pi}{2}) = \parton{1 + X_{(2,1)}(\te+\frac{i\pi}{4})} \parton{1 + X_{(2,1)}(\te-\frac{i\pi}{4})} \parton{1 + X_{(1,1)}} \parton{1 + X_{(3,1)}} \nn \\
\end{split}
\ee
\be
X_{(a,m)}(\te+\frac{i\pi}{4}) X_{(a,m)}(\te-\frac{i\pi}{4}) = \parton{1 + \delta_{m1} \delta_{a2} X_0} \frac{\parton{1 + X_{(a,m+1)}} \parton{1 + X_{(a,m-1)}} }{\parton{1 + \displaystyle\frac{1}{X_{(a+1,m)}}} \parton{1 + \displaystyle \frac{1}{X_{(a-1,m)}} }} \qquad 
\begin{array}{ccc}
&a=1,2,3 \\ &m=1,2,3,...,p-1
\end{array}
\label{o(6)ysys}
\ee
(imposing $X_{(a,0)}=X_{(a,p)}=(X_{(0,m)})^{-1}=(X_{(4,m)})^{-1}=0$ and taking the limit $p\rightarrow\infty$), which may be represented on the diagram in \textit{Fig.}(\ref{o(6)diagram}) and enjoys the usual ({\it uncrossed}) form.

Moving from this $O(6)$ diagram we may think to construct that of \textit{Fig.}(\ref{RSS}) for $N=4, \ka=\infty$ paralleling the graphic folding procedure resulting in the $AdS_4$ digram  \cite{ads4} from that of $AdS_5$, as described previously in the main text. Namely, we can merge together rows $1$ and $3$ in \textit{Fig.}(\ref{o(6)diagram}), while all nodes along the symmetry row $2$ (including the massive node) shall split into two nodes. In particular, the unique massive node $0$ is 'torn' into two, that is, we can imagine, the spinon $0$ and the antispinon $\bar{0}$ in \textit{Fig.}(\ref{RSS}) (for $N=4$). The latter need now to satisfy the 'crossed' equations (\ref{YSS4}).

The physical and mathematical implications of this observation are left for ongoing investigations, also in relation to other folding \cite{fen-ginsp} and quiver \cite{Kuniba:2010ir,Nakanishi:2010tt} procedures.

\end{document}